\documentstyle[12pt,epsf]{article}

\catcode`\@=11 % This allows us to modify PLAIN macros.
\def\lsim{\mathrel{\mathpalette\@versim<}}
\def\gsim{\mathrel{\mathpalette\@versim>}}
\def\@versim#1#2{\vcenter{\offinterlineskip
        \ialign{$\m@th#1\hfil##\hfil$\crcr#2\crcr\sim\crcr } }}
\catcode`\@=12 % at signs are no longer letters

\begin{document}
\begin{titlepage}
  \begin{flushright}
    KUNS-1516 \\[-1mm]
    HE(TH)98/10 \\[-1mm]
    hep-ph/9806400
  \end{flushright}
  \begin{center}
    \vspace*{1.2cm}
    
    {\large\bf Sterile Neutrinos in a Grand Unified Model}
    \vspace{1cm}
    
    {Masako Bando\footnote{E-mail address:
        bando@aichi-u.ac.jp} and
      Koichi Yoshioka\footnote{E-mail address:
        yoshioka@gauge.scphys.kyoto-u.ac.jp}}
    \vspace{7mm}
    
    $^*${\it Aichi University, Aichi 470-02, Japan} \\
    $^{\dagger}${\it Department of Physics, Kyoto University
      Kyoto 606-8502, Japan}
    \vspace{1.5cm}

    \begin{abstract}
      The recent experimental results indicating the neutrino
      oscillations may strongly suggest that at least one more light
      neutrino species is required in order to reconcile the existing
      data. In the simple GUT frameworks, this fact seems difficult to
      preserve the parallelism between quarks and leptons. In this
      letter, we investigate an $SO(10)$ grand unified model with a
      pair of extra generations in addition to the known three
      ones. Using the GUT relations, the obtained neutrino mass matrix
      naturally indicates that one of the $SU(2)_L$ singlet (sterile)
      neutrino is very light and has large mixing with muon neutrino,
      which can explain the atmospheric neutrino anomaly, and the hot
      dark matter neutrino is also provided. The solar neutrino
      problem can be solved by the mixing with muon neutrino
      consistently with quark mixing, namely, the Cabibbo
      angle. \\[5mm]
%      \noindent
%      PACS numbers: 14.60.Pq, 12.10.-g
    \end{abstract}
  \end{center}
\end{titlepage}
\setcounter{footnote}{0}

\newpage

Accumulating data of several experiments have now convinced us that
the neutrinos have non-vanishing masses and mixings. The observed
solar neutrino deficits \cite{Homestake}-\cite{SKAMsol} compared to
the standard solar model calculations \cite{BP} can be explained in
terms of the matter induced resonant oscillation \cite{MSW} with the
oscillation parameters $\Delta m^2 \simeq (0.4-1.1)\times 10^{-5}$
eV$^2$ and $0.003 \lsim \sin^2 2\theta_{ex} \lsim 0.012$
\cite{sol-analy}.\footnote{There is another solution with large mixing
  angle which is less preferable in view of the recent Superkamiokande
  reports on the day-night effect and the electron recoil energy
  spectrum \cite{sol-analy}.} The atmospheric neutrino anomaly
\cite{KAMatm}-\cite{MACRO} also indicates the neutrino oscillation
$\nu_\mu \leftrightarrow \nu_{\tau,s}$ with $\Delta m^2 \sim
10^{-(2-3)}$ eV$^2$ and $0.8 \lsim \sin^2 2\theta_{\mu x} \lsim 1$
\cite{atm-analy}. Another hint of the neutrino masses and mixings
comes from the astrophysics and cosmology. Especially, if the neutrino
is considered as a natural candidate for the hot dark matter component
which is needed to explain the anisotropy of the cosmic microwave
background radiation and so on, it requires the neutrino masses to be
a few eV \cite{DM}. Within the known three neutrino framework, the
only solution which can explain the above experimental results
requires three almost degenerate mass eigenstates with masses $\simeq
O({\rm eV})$ \cite{ADM}. However, it requires fine-tunings or very
hierarchical right-handed neutrino Majorana masses
\cite{GUT3}. Together with the large 2-3 mixing, this is apparently in
contrast to the character of the ordinary quark masses and
mixings. Thus, the simultaneous explanation of the solar, the
atmospheric and the hot dark matter neutrino within the three
generation scenario seems unnatural, in particular within GUT
frameworks \cite{GUTs}. In addition, the accelerator and reactor
experiments also constrain the allowed parameter regions. We shall
comment on these matters later.

One of the natural ways to solve the problem is to introduce extra
neutrinos which must be $SU(2)_L\times U(1)_Y$ singlets (sterile) in
view of the results of the LEP data. Along this way, many theoretical
works are recently investigated \cite{sterile}. However if one
considers that the gauge unification or the left-right symmetry may be
realized in nature, it should be pursued to understand this neutrino
spectrum from the relations in some GUT framework \cite{GUT4}. Then,
the large mixing may originate from the mixing with sterile neutrinos
other than the ordinary three generations since it is expected that
the mixings are small between the ordinary neutrinos.

In this paper we present such an supersymmetric grand unified model
based on the $SO(10)$ gauge group in which an extra light neutrino is
included and naturally has large mixing with the ordinary
neutrinos. In this model, we add a pair of extra vector-like
generations \cite{vector}-\cite{fujikawa} from which a sterile
neutrino arises in addition to the ordinary three ones. The important
feature of the model is that due to the existence of the extra
generations (hereafter, we describe them as 4 and $\bar 4$
generations), all the gauge couplings become asymptotically non-free
while preserving gauge coupling unification \cite{ANFGUT,BSOT}. This
fact yields the strong convergence of Yukawa couplings to their
infrared fixed points (IRFP) \cite{IRFP}, and with this property we
can determine the texture of the quark and lepton mass matrices. In
the previous paper \cite{ANFmass}, we found that the texture is almost
uniquely determined if we impose that the masses of heavy up-type
quarks (top and charm) are realized as their IRFP values. The most
characteristic feature of this texture is that only the second
generation strongly couples to the extra generations. This fact
indicates that the muon neutrino may have a large mixing with the
extra generations which gives the origin of the atmospheric neutrino
anomaly. Moreover, as we shall see later, using the GUT relations for
Yukawa couplings, we can also fix the Majorana mass matrix of the
right-handed neutrinos. Then it is interesting to see how the light
neutrinos can be provided and their mass matrix is predicted in this
$SO(10)$ model.

Before going into the neutrino masses, we first summarize the
ingredients of the previously obtained results which we need to
analyse the neutrino mass matrix. As we have stressed above, in
asymptotically non-free models, the IRFP behaviors can determine the
fate of the low-energy quark Yukawa couplings almost uniquely; all the
quark Yukawa couplings with appreciable strength grow up to be of
order one. So, in the present model, the dominant elements in the
quark mass matrices are of the order of either, the electroweak scale
or the invariant mass scale at which the extra generations are
decoupled (it is expected to be of the order of TeV
\cite{extra-mass}-\cite{STU})\@. Another characteristic feature is the
down to charged lepton mass ratio strongly enhanced by the strong
gauge couplings. It requires that the down and charge lepton sectors,
especially bottom and tau, couple to Higgs fields of $\overline{126}$
representation of $SO(10)$ which induces the ratio 1 : 3 for Yukawa
couplings at the GUT scale. Combined with the enormous QCD enhancement
factor of about $5\sim 6$ (in contrast to $\sim 3$ in the MSSM), it
can correctly reproduce the low-energy experimental value of the
bottom-tau ratio $\sim 1.7$\@. Note that the right-handed neutrino
Majorana masses come from the standard gauge singlet component of
$\overline{126}$-Higgs and therefore may be proportional to the down
and charged lepton sectors.

Since the realistic texture should yield typical hierarchical
structures, we can first fix the leading part of mass matrices
(hereafter for simplicity, $w$ and $M$ are used symbolically to
represent electroweak scale masses and invariant masses of the pair of
the extra generations, respectively). Among the  $5\times 5$ Dirac
mass matrices, it is easily seen that the matrix elements relevant to
the first generation can be neglected because of the hierarchy
structures. Thus, we shall express the mass matrices in $4\times 4$
forms hereafter. The forms of the dominant elements in the quark and
charged lepton mass textures at the GUT scale turn out to be as
follows \cite{ANFmass};
\begin{eqnarray}
  m_u &=& \bordermatrix{
           &  2  &  3  &  4  & \bar 4 \cr
    2 ~    &     &     &  w  &        \cr
    3      &     &  w  &     &        \cr
    4      &  w  &     &     &   M    \cr
    \bar 4 &     &     &  M  &   w    \cr
    }\,,\qquad
  m_d \;=\;
  \bordermatrix{
           &  2  &     3      &  4  & \bar 4 \cr
    2 ~    &     &            &  w  &        \cr
    3      &     & \epsilon w &     &        \cr
    4      &  w  &            &  w  &   M    \cr
    \bar 4 &     &            &  M  &        \cr
    }\,,
  \label{qmass}
  \\
  &&\qquad\quad m_e \;=\; \bordermatrix{
           &  2  &       3      &  4  & \bar 4 \cr
    2 ~    &     &              & 3 w &        \cr
    3      &     & 3 \epsilon w &     &        \cr
    4      & 3 w &              & 3 w &   M    \cr
    \bar 4 &     &              &  M  &        \cr
    }\,.
\end{eqnarray}
The above texture has the following characteristic properties;
(i) The charm quark mass as well as the top quark are determined from
their IRFP values. The charm to top mass ratio is suppressed by the
factor $w^2/M^2$ which comes from the existence of the heavy extra
generations. (ii) It is interesting that the 2-4 (4-2) elements reach
their IRFPs at low energy whose values are of order one. This
indicates that the second generation is strongly coupled to the extra
generations. (iii) The charged lepton masses are reproduced quite
successfully by assuming that the relevant Higgs fields belong to
$\overline{126}$ representation of $SO(10)$ as noted before. (iv) The
$\epsilon$ parameter in the 3-3 elements is needed to reproduce the
correct bottom to strange (or tau to mu) mass ratio and its value is
predicted to be $\sim 0.2$\@. Within this approximation, taking the
parameters as $M_{\rm GUT}\sim 5\times 10^{16}$ GeV, $\alpha_{\rm GUT}
\sim 0.3$ and $\tan \beta \sim 20$, for example, we get the low-energy
predictions at $M_Z$ scale; $m_t \sim 180,\, m_c \sim 1.0,\, m_b \sim
3.1,\, m_s \sim 0.08,\, m_\tau \sim 1.75$ and $m_\mu \sim 0.10\,$ (in
GeV)\@. These are in good agreement with the experimental data
\cite{Mzdata}. The full mass matrices including quark mixing angles
can be obtained by introducing hierarchically very small (less than
the order of $\epsilon^3$) Yukawa couplings. After all, we can get a
reasonable $5\times 5$ GUT-scale texture which explains the
experimental values of the CKM mixing angle. It should be stressed
that the above texture is found to be actually the only possibility
left in view of the IRFP structure.

Let us proceed to the neutrino masses, $m_\nu^D$ (Dirac) and $m_\nu^R$
(right-handed Majorana). Once we fix the texture of quark and charged
lepton, the $SO(10)$ gauge symmetry can relate the neutrino mass
texture to the quark ones. This time we have one more scale of the
right-handed neutrino Majorana mass $M_R$ in addition to $M$ and $w$,
among which a large hierarchy exists; $w < M \ll M_R$.

Now, let us consider the mixing of the first generation which is
responsible for the solar neutrino problem. In the quark sector, it is
known that the 1-2 mixing, that is, the Cabibbo angle is properly
reproduced from the down-quark part only; $\sin \theta_C \simeq
\left(m_d/m_s\right)^{1/2}\sim 0.22$ \cite{oakes}. According to the
GUT relation between quark and lepton Dirac mass matrices, the
corresponding lepton 1-2 mixing angle is $\left(m_e/m_\mu\right)^{1/2}
\sim 0.07$, which is disfavored more than at a $2\sigma$ level for the
MSW small angle solution \cite{FY}. However, the lepton mixing
consists of two parts, the charged lepton and neutrino ones. Since the
GUT relations lead a small mixing in the charged lepton sector, the
large mixing angle ($\sin \theta \sim 1/\sqrt{2}$) of the second
generation required by the recent Superkamiokande report should come
from the neutrino side in the present model. Then the lepton 1-2
mixing is predicted that $\sin \theta_{e\mu} \sim
\left(m_e/m_\mu\right)^{1/2}\times 1/\sqrt{2} \sim 0.05$ which is now
well within the desired range for the solar neutrino problem. After
all, we do not have to consider the mixing of the first generation
neutrino with the other ones, if only the second generation neutrino
mixes strongly with the other generations except for the first one
\cite{BKY}. It is noted that from the Superkamiokande atmospheric
neutrino data (the zenith angle distribution of the $e$-like and
$\mu$-like events data) and the recent results of the CHOOZ
long-baseline oscillation experiment \cite{chooz}, the large angle
$\nu_e \leftrightarrow \nu_\mu$ oscillation is found to be disfavored
for the solution to the atmospheric neutrino anomaly
\cite{atm-analy}. So, the above mechanism seems to work naturally and
to be a likely scenario in GUT models. In the following, therefore, we
can consider the $4\times 4$ neutrino mass matrices. From the quark
texture (\ref{qmass}), we can get the following texture for neutrinos;
\begin{eqnarray}
  m_\nu^D \;=\; \bordermatrix{
           &  2  &  3  &  4  & \bar 4 \cr
    2 ~    &     &     &  w  &        \cr
    3      &     &  w  &     &        \cr
    4      &  w  &     &     &   M    \cr
    \bar 4 &     &     &  M  &   w    \cr
    }\,,&&
  m_\nu^R \;=\;
  \bordermatrix{
           &  2  &       3      &  4  & \bar 4~ \cr
    2 ~    &     &              & M_R &         \cr
    3      &     & \epsilon M_R &     &         \cr
    4      & M_R &              & M_R &         \cr
    \bar 4 &     &              &     &         \cr
    }\,,
  \label{numass}
\end{eqnarray}
where we use the GUT relation $m_\nu^D=m_u$ and the fact that
$m_\nu^R$ comes from the $\overline{126}$-Higgs fields, namely,
$m_\nu^R\propto m_d \,(m_e)$\@. The above neutrino texture indicates
that; (i) One extra (sterile) neutrino in the $\bar 4$ generation is
left to be almost massless and may couple strongly to the second
generation (muon) neutrino. (ii) The third generation right-handed
Majorana mass is a little smaller than the others. This yields a
heavier left-handed tau neutrino which can be the hot dark matter
component. In the above texture we have assumed that the up-type
quarks as well as neutrinos couple to 10-Higgs and especially, the
$\bar 4$-$\bar 4$ elements do not come from 126-Higgs (not
$\overline{126}$). This may be easily realized when one introduces
relevant Higgs multiplets with a flavor $U(1)$ (gauge) symmetry (see
the appendix). However, it is interesting that almost all parts of the
above texture can be fixed from the characteristic IR property of this
model without such any symmetry arguments.

As seen from the textures (\ref{qmass})--(\ref{numass}), the third
generation is almost decoupled and can be neglected in the following
analyses. In the remaining part, two of six neutrinos (the second and
fourth right-handed neutrinos) are of the order of the intermediate
scale $M_R$\@. In this way the neutrino texture is reduced to $4\times
4$ matrix with light elements. Then, the problem is whether the mixing
angle between light neutrinos can become very large. After integrating
out the heavy right-handed neutrinos of the second and fourth
generations, we get the following mass matrix in the basis of
$(\nu_{2_2}, \nu_{\bar 4_1}, \nu_{4_2}, \nu_{\bar 4_2})$ (the second
subscripts represent the transformation properties under the
$SU(2)_L$),
\begin{eqnarray}
  \pmatrix{
    2\alpha m & \alpha m' &  m  &     \cr
    \alpha m' &           &  m' &  w  \cr
       m      &     m'    & -m  &  M  \cr
              &     w     &  M  &
    },
  \label{matrix1}
\end{eqnarray}
where $m$ and $m'$ are masses induced by seesaw mechanism
\cite{seesaw} $(m\sim\frac{w^2}{M_R}\,,\, m'\sim\frac{w M}{M_R})$ and
are much smaller than $w$ and $M$\@. Therefore we are left with two
very light neutrinos with masses $\sim O(m,m')$ which mainly come from
$\nu_{2_2}$ and $\nu_{\bar 4_1}$\@. In the above matrix, $M$ is an
invariant mass of the extra lepton doublets. Its range is estimated as
$M\gsim 200$ GeV if one takes account of the constraints for the extra
vector-like quark masses $(\gsim 1 \mbox{ TeV})$ from the FCNC
\cite{fujikawa} and $S,T$ and $U$ parameters \cite{STU}, and the
relative QCD enhancement factor $(\sim 5)$ between quarks and leptons
in this model. There also appear non-zero matrix elements with a
factor $\alpha$ which come from the induced neutrino Dirac mass
elements via one-loop renormalization group. This $\alpha$,
representing the ratio of induced to tree-level Dirac masses, is
almost independent of the input parameters ($\tan \beta, \alpha_{\rm
  GUT}$, etc.)\@ and its typical value is $|\alpha| \sim 0.1$\@. By
diagonalizing the mass matrix (\ref{matrix1}), the mixing angle
between the light neutrinos $(\nu_{2_2}, \nu_{\bar 4_1})$ becomes,
\begin{eqnarray}
  \tan 2\theta &=& \frac{2m'\alpha\cos \phi - 2m\sin\phi}
  {m'\sin 2\phi +m(2\alpha+\sin^2\phi)}\,, \\[1mm]
  && \tan\phi \>\equiv\> \frac{w}{M}\,. \nonumber
\end{eqnarray}
Since $\,m/m',\, \tan \phi \sim w/M \ll 1$, we have,
\begin{eqnarray}
  \tan 2 \theta &\sim& \frac{\alpha}{\sin \phi}\,.
\end{eqnarray}
By taking the typical values of $\alpha$ and $w$, the mixing angle
becomes,
\begin{eqnarray}
  \sin^2 2\theta &\sim&
  \frac{1}{1+\displaystyle{\left(\frac{350}{M\;
          \mbox{(GeV)}}\right)^2}\!\cos^2 \beta}\,,
\end{eqnarray}
with $\tan \beta$, a ratio of the vacuum expectation values of two
doublet Higgses. From this, for $\tan \beta \gsim 3$, we can naturally
get the large mixing angle for suitable parameter range ($M \gsim 200$
GeV) (Figure 1)\@.
\begin{figure}[htbp]
  \begin{center}
    \epsfxsize=8cm \ \epsfbox{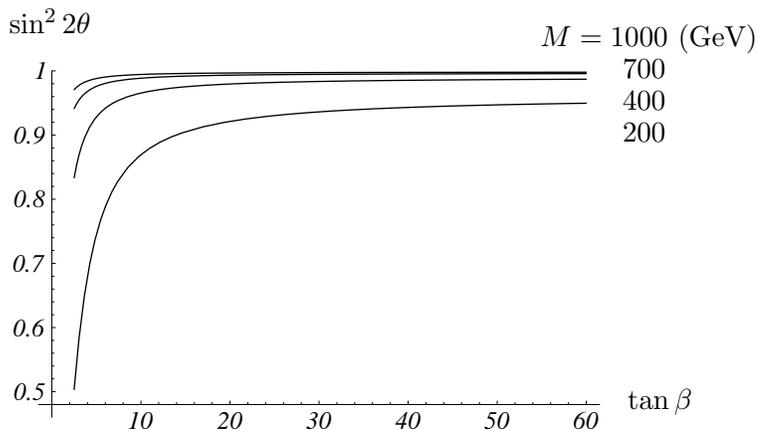}
    \put(10,13){\small $\tan \beta$}
    \put(-224,155){\small $\sin^2 2\theta$}
    \put(-23,150){\small $M=1000$ (GeV)}
    \put(8,138){\small $700$}
    \put(8,126){\small $400$}
    \put(8,114){\small $200$}
    \caption{The mixing angle between the second and anti-fourth
      generations}
  \end{center}
\end{figure}

To be more exact, three blanks except for the right-bottom element in
the matrix (\ref{matrix1}) are radiatively induced as well if the
invariant masses come from Yukawa couplings to a singlet field
\cite{ANFmass}. Then the light neutrino mass matrix becomes,
\begin{eqnarray}
  \pmatrix{
    2\alpha m & \alpha m'   &  m  & \gamma M \cr
    \alpha m' & \alpha' m'' &  m' &    w     \cr
        m     &      m'     & -m  &    M     \cr
    \gamma M  &      w      &  M  & 
    }\,,
  \label{matrix2}
\end{eqnarray}
where $m''$ represents seesaw induced mass $(m''\sim
\frac{M^2}{M_R})$, and $\alpha'$ and $\gamma$ are relative ratios of
the renormalization group induced mass parameters to the tree level
ones. They are again almost independent of the input values. The
typical values are $|\alpha'| \sim 0.01$ and $|\gamma| \sim
0.1$\@. This texture (\ref{matrix2}) is just a realization of the
recently proposed so-called singular seesaw matrix \cite{singular},
and two out of the above four neutrinos remain very light. An analytic
expression for the mixing angle of the remaining two neutrinos is,
\begin{eqnarray}
  \tan 2\theta &=& 2\Bigl( -m''\alpha'\cos\phi\cos\phi' +m'(\sin\phi
  \sin\phi'+\alpha \cos\phi'\cos 2\phi) \Bigr. \nonumber \\
  &&\Bigl.\hspace*{3mm} -m\cos\phi(\sin\phi' -2\alpha\sin\phi
  \cos\phi') \Bigr) \Big/ \Bigl
  ( m''\alpha'(\sin^2\phi+\cos^2\phi\cos^2\phi') \Bigr. \nonumber\\
  &&\Bigl.\hspace*{6mm} +m'\bigl(\cos\phi\sin 2\phi' -\alpha\sin 2\phi
  (1+\cos^2\phi') \bigr) \Bigr. \nonumber \\
  &&\Bigl.\hspace*{9mm} +m\,(\sin^2\phi' +\sin\phi \sin 2\phi'
  +2\alpha (\cos^2\phi-\sin^2\phi \cos^2\phi')\Bigr), \\[1mm]
  &&\qquad\qquad \tan \phi \;=\; \frac{\gamma M}{w},\quad\;\; \tan
  \phi' \;=\; \frac{\gamma}{\sin\phi}\,.
\end{eqnarray}

Now for the numerical estimations. Since the third generation neutrino 
is identified to the hot dark matter component and it is almost
decoupled from the other generations, the intermediate scale $M_R$ is
mainly determined from the eigenvalue $m_3$\@. We find that the
desired tau neutrino mass is obtained if we take $M_R$ as $10^{12}\;
\mbox{GeV} \lsim M_R \lsim 10^{13}$ GeV (Figure 2)\@. Then, for the
solar and atmospheric neutrino anomalies, the $\Delta m^2$ and the
mixing angles depend on the other parameters and especially are
sensitive to $\tan \beta$ and $M$ as indicated above. In Figure 3--5,
we display acceptable solutions as an example and typical values of
the masses and mixing angles are,
\begin{eqnarray}
  \Delta m_{12}^2 &\simeq& 1.0\times 10^{-5} \mbox{ eV}^2\,,\quad\;
  \sin^2 2\theta_{e\mu} \;\simeq\; 0.012\,, \\
  \Delta m_{2\bar 4}^2 &\simeq& 1.1\times 10^{-3} \mbox
  { eV}^2\,,\quad\; \sin^2 2\theta_{\mu s} \;\simeq\; 0.82\,, \\
  m_3 &\simeq& \mbox{a few eV}\,,
\end{eqnarray}
for $M_R\sim 4\times 10^{12}$ GeV, $\tan \beta \sim 30\,$ and $M\sim
250$ GeV\@. These are in good agreement with the experimental
observations of the solar, atmospheric and hot dark matter neutrinos.
\begin{figure}[htbp]
  \parbox{6.6cm}{
    \epsfxsize=6.5cm \ \epsfbox{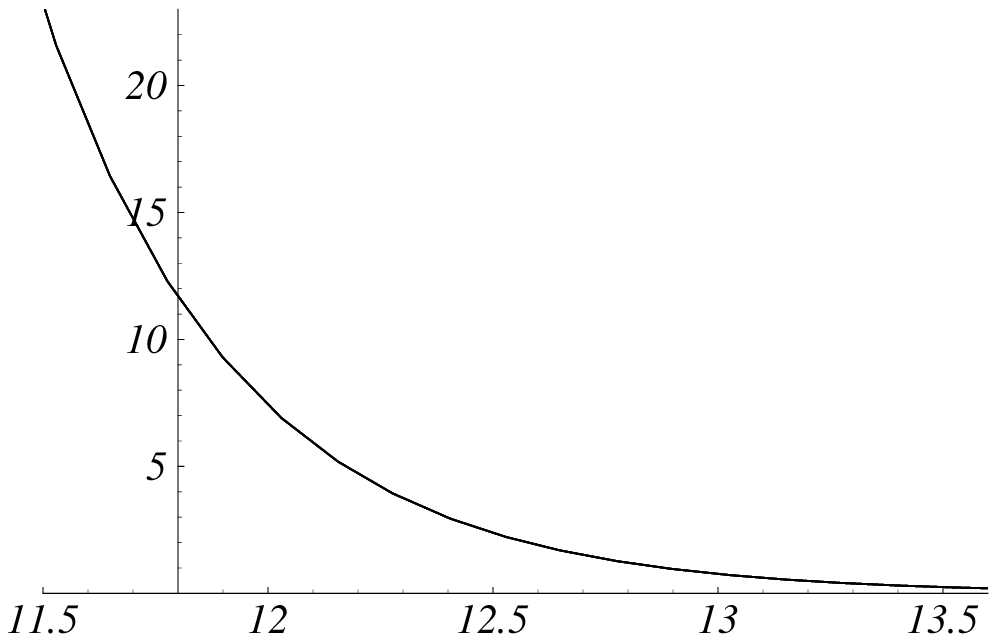}
    \put(5,10){\footnotesize $\log_{10} M_R$}
    \put(-165,128){\footnotesize $m_3$ (eV)}
    \caption{The $M_R$ dependence of the eigenvalue $m_3$ (mass of the
      hot dark matter neutrino)}
    }
  \hspace*{1.5cm}
  \parbox{6.6cm}{
    \epsfxsize=6.5cm \ \epsfbox{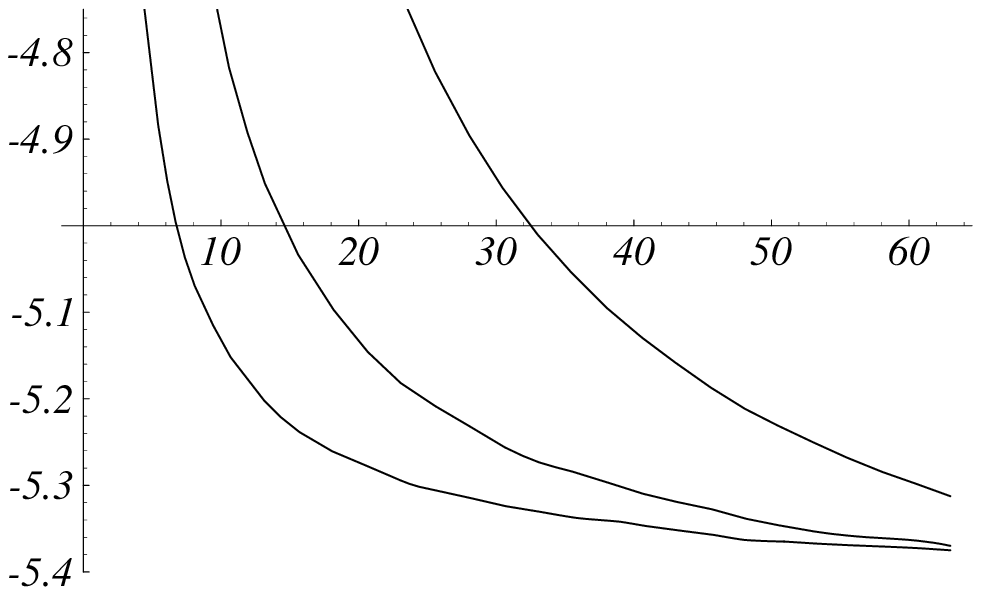}
    \put(5,73){\footnotesize $\tan \beta$}
    \put(-183,124){\footnotesize $\log_{10} \Delta m^2$}
    \put(-18,37){\footnotesize $M=200$ (GeV)}
    \put(6,18){\footnotesize $400$}
    \put(6,8){\footnotesize $600$}
    \caption{The predicted value of $\Delta m^2$ for the solar
      neutrino anomaly.}
    }
\end{figure}

\begin{figure}[htbp]
  \parbox{6.6cm}{
    \epsfxsize=6.5cm \ \epsfbox{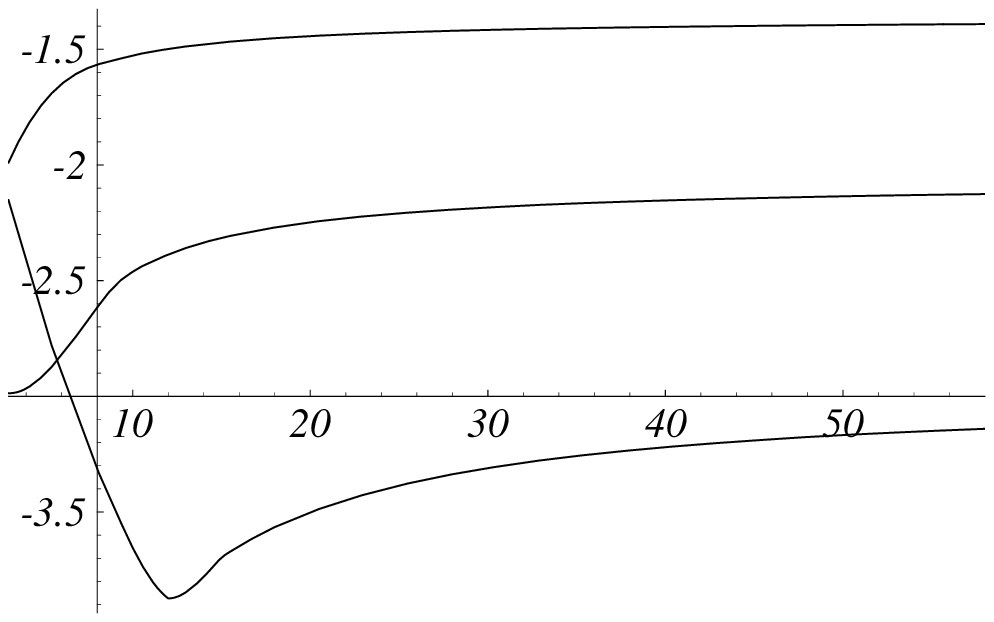}
    \put(6,43){\footnotesize $\tan \beta$}
    \put(-183,126){\footnotesize $\log_{10}\Delta m^2$}
    \put(-43,123){\footnotesize $M=600$ (GeV)}
    \put(-20,70){\footnotesize $400$}
    \put(-20,25){\footnotesize $200$}
    \caption{The predicted value of $\Delta m^2$ for the atmospheric
      neutrino anomaly.}
    }
  \hspace*{1.5cm}
  \parbox{6.6cm}{
    \epsfxsize=6.5cm \ \epsfbox{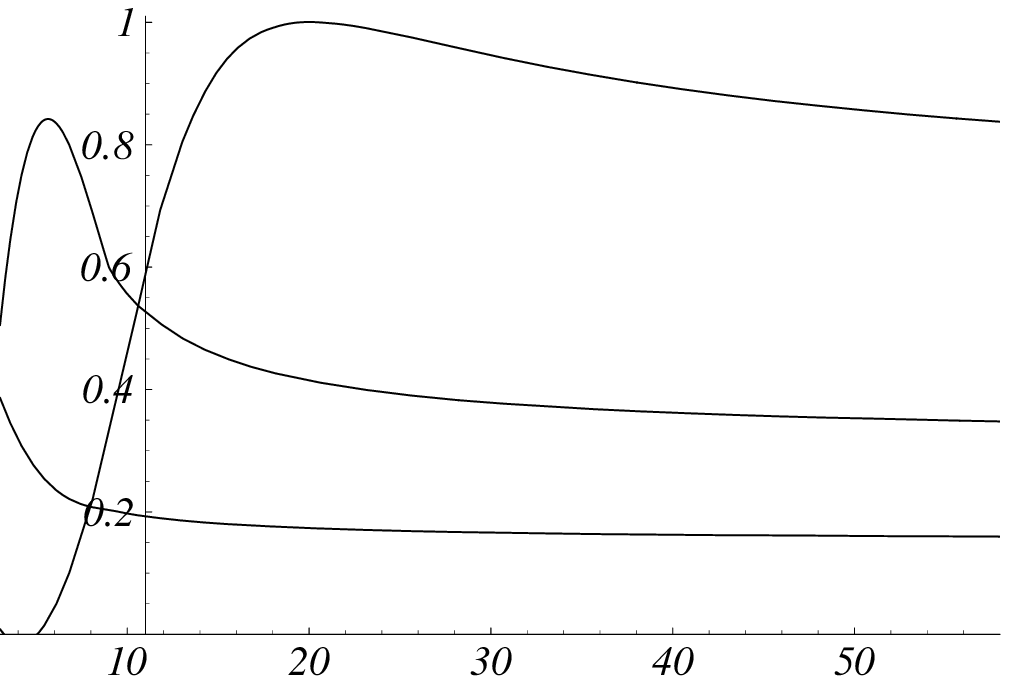}
    \put(5,5){\footnotesize $\tan \beta$}
    \put(-171,130){\footnotesize $\sin^2 2\theta$}
    \put(-37,110){\footnotesize $M=200$ (GeV)}
    \put(-13,50){\footnotesize $400$}
    \put(-13,30){\footnotesize $600$}
    \caption{The predicted value of $\sin^2 2\theta$ for the
      atmospheric neutrino anomaly.}
    } 
\end{figure}

A few comments are in order concerning the other experimental
results. In this model, the sterile neutrino has a large mixing with
the muon neutrino to solve the atmospheric neutrino anomaly. In this
scheme, the positive LSND results of the $\nu_e\leftrightarrow
\nu_\mu$ oscillation \cite{LSND} can be reconciled at a $3\sigma$
level only \cite{4analy} with the indirect oscillation \cite{indirect}
through the tau neutrino. Or, it can be certainly explained by the
sterile neutrino with a heavier mass but at this time the zenith angle
dependence of the atmospheric neutrino data is not expected. On the
other hand, the recent results of the KARMEN experiment \cite{Karmen}
seems to exclude almost all the allowed parameter region of the LSND,
so it may not be necessary to take the LSND results seriously in this
paper. The discrimination between two oscillation scenarios,
$\nu_\mu\leftrightarrow\nu_\tau$ and $\nu_\mu\leftrightarrow\nu_s$,
for the solution to the atmospheric neutrino anomaly will be made by
the ongoing and forthcoming experiments observing various quantities
\cite{disc}. The recent Superkamiokande reports indicate that the
observed suppression of the $NC$ induced $\pi^0$ events is consistent
with $\nu_\mu\leftrightarrow\nu_\tau$ oscillation but they have not
excluded $\nu_\mu\leftrightarrow\nu_s$ oscillation as yet. The
cosmological and astrophysical implications in the existence of the
fourth light neutrino should also be addressed, especially, the
big-bang nucleosynthesis scenario which severely constrains the
effective number of light neutrino species, or equivalently the mixing
between the active and sterile neutrinos. However, according to the
recent estimations \cite{number}, more than four light neutrinos are
acceptable and there is no constraint on the mixing angles. Even if
the constraint is revalued and the allowed number turns out to be less
than four, there is an interesting and simple mechanism which has
recently been proposed \cite{leptonasym}. In order to avoid the
constraints, it requires the large lepton asymmetry $(\gsim 10^{-5})$
for which a small mixing between the active (tau) and sterile
neutrinos is needed. This can be easily realized in the present
model.

In summary, we have investigated a supersymmetric $SO(10)$ model with
a pair of extra vector-like generations. In this model, the textures
are almost uniquely determined by the IRFP structures due to the
asymptotically non-freedom of gauge couplings, and the GUT relations
between quark and lepton. We have particularly examined the neutrino
sector and found that; (i) By assuming that the $\bar 4$ generation
couples to 10-Higgs, one of the extra $SU(2)_L$ singlet neutrino is
made to be very light which comes into play as a sterile neutrino, and
this neutrino has very large mixing with the muon neutrino which can
explain the atmospheric neutrino anomaly. (ii) The texture requires
that the third generation right-handed neutrino is a little lighter
than the others, resulting in the heavier left-handed tau neutrino to
reach to the hot dark matter candidate. (iii) The solar neutrino
problem can be explained by the mixing with muon neutrino,
consistently with the mixing angle expected from the GUT relation with
the Cabibbo angle.

Noting that the supersymmetry breaking scale is of the same order as
the invariant masses of the extra generations, we may discover the
extra fermions when supersymmetry is found. Moreover, by muon
colliders \cite{muon} the extra generations may be explored easily
since in the present model the second generation strongly couples to
the extra ones. It is interesting that the extra generations appear
themselves via the second generation in the neutrino sector. We would
like to also stress that neutrinos are more appropriate subjects to be
investigated to seek for the extra generations, and hope that the
sterile neutrino scenario will be confirmed by the experiments of new
generation.

\subsection*{Acknowledgements}

We would like to thank T.\ Yanagida for stimulating discussions while
the Summer Institute '97 held at Yukawa Institute in August
1997. Discussions with J.\ Sato and N.\ Maekawa are also
appreciated. M.\ B. is supported in part by the Grant-in-Aid for
Scientific Research from Ministry of Education, Science and
Culture and K.\ Y. by the Grant-in-Aid for JSPS Research fellow. 

\vspace*{2mm}

\subsection*{Appendix}

The texture zeros can arise due to symmetries in the underlying string
or GUT theory. In this appendix, we show an example which reproduces
the textures adopted in this paper. Although there may be many
possibilities that realize the desired texture and among them there
might exist simpler choices, it would be instructive to see how the
desired patterns of the texture come about from such kind of flavor
symmetries.

Let us consider the case in which the matter and Higgs fields having
additional flavor $U(1)$ charges. We consider the following Higgs
multiplets of $SO(10)$ representation; $\Phi_{1,2}(210)$,
$\Delta_{1,2}(126)$, $\bar \Delta_{1,2}(\overline{126})$,
$H_{1,2,3,4}(10)$, $\theta(1)$ as well as the matter superfields
$\Psi_{1,2,3,4}(16)$ and $\bar \Psi_{\bar 4}(\overline{16})$\@. Their
charges under the $U(1)$ symmetry are given in Table 1\@. Then, the
gauge and flavor invariant superpotential becomes;
\begin{eqnarray}
  W &=& (H_1+\bar \Delta_1) \Psi_2\Psi_4 + H_2 \Psi_3 \Psi_3 + \bar
  \Delta_1 \theta \Psi_3 \Psi_3 +\bar \Delta_2 \Psi_4 \Psi_4 +H_3
  \bar \Psi_{\bar 4} \bar\Psi_{\bar 4} \nonumber \\ 
  &&\qquad +H_1 \bar \Delta_1 \Phi_2 +H_3 \Delta_1
  \Phi_1 + W_m + W_G.
\end{eqnarray}
The term $W_m$ contains the relevant mass terms of the above Higgs
fields by some of which the $U(1)$ flavor symmetry may be softly
broken. Suppose that $SO(10)$ gauge symmetry is broken down to the
standard gauge group by $W_G$ for appropriate choice of Higgs
couplings (probably, including more Higgs multiplets (45-, 54-Higgs)
in addition to the above ones)\@. The vacuum expectation values of
singlet components in $\Phi$'s can break not only the $SO(10)$ but
also D-parity \cite{D-parity}. This parity breaking is favored by
several phenomenological reasons \cite{D-break} and especially it can
suppresses direct left-handed neutrino Majorana mass terms \cite{C-M}
which we do not consider in this paper. As is easily seen, since all
the desired Yukawa couplings are contained in the above
superpotential, we must include the terms so that one linear
combination of the doublet Higgses may remains light in $W_m$ (and
$W_G$) \cite{DT}. This can be easily done by the choice of the softly
broken mass terms in $W_m$, for example;
\begin{eqnarray}
  W_m &=& m_1H_1H_4 + m_2H_2H_4 + m_3\Delta_1\bar\Delta_2 +
  m_4\Delta_2\bar\Delta_1 + m_5\Delta_2\bar \Delta_2\,.
\end{eqnarray}
With these terms together with the other ones in $W$, a pair of linear
combinations of $H_1,H_2$ (for up-type doublet Higgs) and $H_3,\bar
\Delta_1,\bar \Delta_2$ (for down-type one) remain light in the
low-energy region and give mass terms to the matter superfields,
provided that the phenomenologically favored breaking chain
\cite{chain} is supposed.
\vspace*{5mm}

\begin{center}
  \begin{tabular}{|c|c|c|c||c|c|c|c|c|c|c|c|c|c|c|} \hline
    $\Psi_2$ & $\Psi_3$ & $\Psi_4$ & $\bar \Psi_{\bar 4}$ & $H_1$ &
    $H_2$ & $H_3$ & $H_4$ & $\Delta_1$ & $\Delta_2$ & $\bar
    \Delta_1$ & $\bar \Delta_2$ & $\Phi_1$ & $\Phi_2$ & $\theta$ \\
    \hline\hline
    3 & 1 & 0 & $-2$ & $-3$ & $-2$ & 4 & 1 & $-2$ & $-6$ & $-3$ & 0
    & $-2$ & 6 & 1 \\ \hline
  \end{tabular} \\[4mm]
  Table 1: $U(1)$ quantum number assignments
\end{center}

\vspace*{1cm}
%\newpage

\setlength{\baselineskip}{16pt}

\end{document}